\documentclass[onecolumn,showpacs,superscriptaddress,amsmath,amssymb]{revtex4}

\usepackage{graphicx}

\usepackage{amssymb}

\usepackage{bm}
\usepackage{times}

\newcommand{\stl}[1]{\mbox{$ \hspace{0.1em}
      \stackrel{\rule{0.4pt}{0.275ex}\hspace{0.40em} \!\!\!
      \overline{\hspace{0.06em}\vphantom{\rule{0.4pt}{0.0ex}}
      \hphantom{\mbox{$\displaystyle #1$}}
      \hspace{0.06em} } \!\!\!\hspace{0.40em}\rule{0.4pt}{0.275ex}}
      {#1}\hspace{0.2em}$}}
\newcommand{\ave}[1]{\left\langle#1\right\rangle}

\newcommand{\Ave}[1]{\left\langle\!\left\langle{#1}\right\rangle\!\right\rangle}
\newcommand{\aveqe}[1]{\left\langle#1\right\rangle_\Lambda}

\newcommand{\kb}{k_{\rm B}}
\newcommand{\QE}{\psi^\ast}
\newcommand{\psiQE}{\psi^\ast}
\newcommand{\ba}{\mathbf{a}}
\newcommand{\bB}{\mathbf{B}}
\newcommand{\bC}{\mathbf{C}}
\newcommand{\bF}{\mathbf{F}}
\newcommand{\bG}{\mathbf{G}}
\newcommand{\bM}{\mathbf{M}}

\newcommand{\bu}{\mathbf{u}}

\newcommand{\bP}{\mathbf{P}}

\newcommand{\bzero}{\mathbf{0}}
\newcommand{\bone}{\mathbf{1}}
\newcommand{\bkappa}{\mbox{\boldmath${\kappa}$}}
\newcommand{\dgamma}{\mbox{\boldmath${\dot{\gamma}}$}}
\newcommand{\bLambda}{\mbox{\boldmath${\Lambda}$}}
\newcommand{\bTheta}{\mbox{\boldmath${\Theta}$}}
\newcommand{\btau}{\mbox{\boldmath${\tau}$}}

\newcommand{\setM}{\underline{M}}
\newcommand{\setL}{\underline{\Lambda}}

\newcommand{\cRu}{\hat{{R}}_{\bu}}
\newcommand{\DD}{\overline{D}_r}
\newcommand{\A}[1]{\stl{\ba_{#1}}}
\newcommand{\La}[1]{\stl{\bLambda_{#1}}}
\newcommand{\Afourqe}{\ba_4^\ast}
\newcommand{\OP}{\stl{\ba_2}}

\begin{document}



\title{Canonical Distribution Functions in Polymer Dynamics: 
II. Liquid--Crystalline Polymers}

\newcommand{\TUB}{Institut\ f\"ur\ Theoretische\ Physik, 
Technische Universit\"at Berlin,
Hardenbergstr.~36, D-10623 Berlin, Germany}

\newcommand{\ETH}{ETH
Z\"urich, Department of Materials, Institute of Polymers, CH-8092
Z\"urich, Switzerland}

\author{Patrick Ilg}
\email[]{ilg@physik.tu-berlin.de}
\affiliation{\TUB}

\author{Iliya V. Karlin}
\affiliation{\ETH}

\author{Martin Kr{\"o}ger}
\affiliation{\TUB, \ETH}

\author{Hans Christian {\"O}ttinger}
\affiliation{\ETH}

\date{\today}





\begin{abstract}
The quasi--equilibrium approximation is employed 
as a systematic tool for solving the problem of 
deriving 
constitutive equations 
from 
kinetic models of liquid--crystalline 
polymers. 
It is demonstrated how kinetic models of liquid--crystalline polymers 
can be approximated in a systematic way, 
how canonical distribution functions can be derived from the 
maximum entropy principle and how constitutive equations are derived 
therefrom. 
The numerical implementation of the constitutive equations 
based on the intrinsic dual structure of the quasi--equilibrium 
manifold thus derived 
is developed and 
illustrated 
for particular examples. 
Finally, a measure of the accuracy of the quasi--equilibrium approximation 
is proposed that can be implemented into the numerical integration of 
the constitutive equation. 
\end{abstract}



\maketitle

\section{Introduction}
\label{intro}
This paper continues a systematic approach to the derivation and 
numerical implementation of constitutive equations for complex 
fluids, initiated in Ref.~\cite{IKO02}. 
In this approach, the systematic application of the quasi--equilibrium 
approximation to kinetic models is proposed, including the derivation
of canonical distribution functions (CDF) and their use to 
obtain closed form constitutive equations. 
Also the natural numerical implementation of the resulting constitutive 
equations based on the dual structure of the quasi--equilibria 
is discussed together with a measure of the accuracy 
of the quasi--equilibrium approximation. 
While in Ref.~\cite{IKO02} dilute solutions of flexible polymers were 
considered, the present paper extends this study to  
kinetic models of liquid--crystalline polymers. 
Peculiarities due to the mean--field nature of models for 
liquid crystalline polymers are discussed. 
A more detailed presentation of the subject can be found in 
\cite{Ilg01}. 

Liquid--crystalline polymers have attracted considerable attention 
due to their capability of forming a highly oriented phase 
\cite{dGE74}. 
In polymer processing, for example, 
it is very important to understand the interplay between the 
the tendency towards orientational ordering and the 
orientational effects due to the flow field \cite{Bu98,BIRD87}. 
Kinetic models of liquid--crystalline 
(rigid rodlike) 
polymers were introduced 
by Hess and Doi \cite{He76a,He76b,Doi81} and are now used by many authors. 
Kinetic models of the dynamics of liquid--crystalline polymers 
are able to reproduce qualitatively most of the experimental results 
for the steady state in homogeneous flows \cite{DoiEd86}. 
Predictions of the kinetic models for transient viscoelastic phenomena 
are in general less reliable \cite{Wis81}. 
The main limitation of the kinetic model responsible for this failure is 
probably the neglect of defects of the orientational ordering. 
Recently, some work has been done in order to implement spatial variations 
of the orientational order into the kinetic model \cite{KrSe92,Gio00}. 
For clarity, we restrict ourselves here to spatially homogeneous systems, 
and the corresponding kinetic models. 
The present approach can be easily extended to include also 
spatially inhomogeneous systems.

Numerical implementation of 
kinetic models of liquid--crystalline polymers 
in direct numerical flow calculations 
is in general very computationally intensive. 
However, 
kinetic models of polymer dynamics may serve as a starting point 
for the derivation of constitutive equations. 
In general, this derivation is not straightforward but requires
approximations to the underlying kinetic model. 
The need for so--called closure approximations occurs in many branches of 
statistical physics and several suggestions for such approximations 
have been proposed in the literature 
(see e.g.~\cite{BALIAN92} and references therein). 
A particularly successful approach for the derivation of constitutive 
equation is the so--called method of invariant manifold \cite{GoKa94}. 
This method identifies relevant manifolds for the reduced description 
from which the constitutive equations can be derived. 
For example, it was found in Ref.~\cite{ZmKaDe00} with the help of
this method that the universal limit in the dynamics of dilute 
polymer solutions is given by the (revised) Oldroyd 8--constant model.  

In Ref.~\cite{IKO02} and in the present work, the quasi--equilibrium 
approximation is employed in order to derive a manifold formed 
by a set of canonical distribution functions (CDF). 
The quasi--equilibrium approximation is used successfully also 
in other branches of statistical physics, 
like for example chemical reaction kinetics, plasma and ferrofluids 
(see also Refs.~\cite{BALIAN92,GORBAN84}). 
This approximation to the 
dynamics shows several desirable features like conservation of the 
dissipative nature of the dynamics, conservation of positive--definiteness 
of distribution functions and a thermodynamic structure including 
dual variables.  
Improvements on this approximation can be found by either including more 
variables into the QEA, by constructing improved manifolds according 
to the method of invariant manifold proposed in Ref.~\cite{GoKa94} or 
by using a combined micro--macro simulation technique proposed in  
Ref.~\cite{GoKaIlOe01}.
Here, we show how the quasi--equilibrium approximation can be used 
in order to obtain closed constitutive equations for liquid--crystalline 
polymers. We also present 
an algorithm for the numerical implementation of the constitutive equations. 
In addition, a measure for the accuracy of the approximation is 
suggested that does not require solutions to the kinetic model and 
can therefore be used while solving the constitutive equation. 

Recently, 
the idea of using a set of canonical distribution functions (CDF), 
which are postulated and not based on the quasi--equilibrium approximation,
for obtaining closures to kinetic models of dilute solutions of 
flexible polymers has also been used in Refs.~\cite{LiHaJaKeLe98,LiKeLe99}.

This paper is organized as follows. 
Kinetic models of the dynamics of liquid--crystalline polymers (LCPs) 
are reviewed in Sec.~\ref{models}. 
Canonical distribution functions and their use in the description 
of the dynamics of LCPs is described in Sec.~\ref{QEA}. 
Closed form constitutive equations are derived using CDF. 
Special emphasize is paid to so--called alignment tensor models. 
In Sec.~\ref{defect}, a measure for the accuracy of the approximate 
description of polymer dynamics with CDF is proposed. 
In Sec.~\ref{dualLCP}, 
some numerical results in steady shear flow are presented. 
Finally, some conclusions are offered in Sec.~\ref{end}.

\section{Kinetic Models of Liquid--Crystalline Polymers}
\label{models}
Consider a solution of liquid--crystalline polymers composed of 
rigid rodlike polymeric molecules. 
The kinetic models describe the microstate of liquid--crystalline polymers 
by the one--particle distribution function $\psi(\bu;t)$, 
the probability density  
of finding a particle oriented along the unit vector $\bu$ at time 
$t$. 
For simplicity, only spatially homogeneous systems are 
considered. 

\subsection{Kinetic Equations of Polymer Dynamics}
Equations of motions for rigid rod models were proposed 
by Hess and Doi, \cite{He76a,He76b,Doi81}, and can be found in 
most of the textbooks on polymer kinetic theory 
(see e.g.~\cite{dGE74,DoiEd86,BIRD87}). 
In the presence of a given homogeneous velocity gradient 
$\bkappa$, 
the time evolution of $\psi$ may be written as 
\begin{equation} \label{Doi_kin}
        \partial_t\psi = -\cRu\cdot\left[ 
        \bu\times\left( \bkappa\cdot\bu\psi \right) \right] +  
        \cRu\cdot\hat D_{\rm r}\psi\cRu 
        \left(\frac{\delta F}{\delta \psi(\bu)} \right).
\end{equation}
Here, $\cRu=\bu\times\partial/\partial \bu$ is the rotational operator, 
$\partial/\partial\bu$ the gradient on the unit sphere, 
$\hat D_{\rm r}$ the rotational diffusivity and  
$\delta/\delta\psi$ the Volterra functional derivative. 
The dimensionless free energy functional per molecule, 
$F[\psi]=F_0[\psi]+F_1[\psi]$, 
can be split into the entropy of molecular alignment, 
\begin{equation} \label{F0_def}
        F_0[\psi] = \int\!d\bu\, \psi(\bu)\ln\psi(\bu),
\end{equation}
and the free energy contributions of interactions $F_1$. 
In the second virial approximation, $F_1$ is given by 
\begin{equation} \label{F1_def}
        F_1[\psi] = \frac{\nu}{2} \Ave{\beta(\bu,\bu')}, 
\end{equation}
where $\nu=n_{\rm p}d\ell^2$ denotes the reduced excluded volume 
of $n_{\rm p}$ rods of length $\ell$ and diameter
$d$ per unit volume. 
Here and below we use the following notations for averages: 
$\ave{f(\bu)}\equiv\int\!d\bu f(\bu)\psi(\bu)$,
$\Ave{f(\bu,\bu')}\equiv\int\!\!\int\! d\bu d\bu'f(\bu,\bu')\psi(\bu)\psi(\bu')$, 
where integration is performed over the three--dimensional unit 
sphere. 

The kinetic equation (\ref{Doi_kin}) is supplemented by the expression for 
the polymer contribution to the stress tensor 
\cite{DoiEd86,IKOe99}, 
\begin{equation} \label{stress_LCP}
        \btau^{\rm p}[\psi] 
        = -3n_{\rm p}\kb T
        \ave{[\bu\times\cRu \frac{\delta F}{\delta\psi(\bu)}] \bu}.
\end{equation}
In Eq.~(\ref{stress_LCP}), viscous contributions to $\btau^{\rm p}$ 
are neglected since they are generally assumed to be negligible in 
liquid--crystalline polymers \cite{DoiEd86}.  
When Eq.~(\ref{stress_LCP}) is used in the Navier--Stokes equations, 
the latter, together with the kinetic equation (\ref{Doi_kin}), 
constitute a closed system describing the dynamics of the solution.

The configuration dependent diffusion coefficient $\hat{D}_{\rm r}$
describes the hindrance of rotations due to neighboring rods. 
Following Doi and Edwards \cite{DoiEd86}, $\hat{D}_{\rm r}$
is approximated by
\begin{equation} \label{Dr_approx}
        \hat D_{\rm r} \approx \overline D_{\rm  r} = 
        D_{\rm r}\left[\frac{4}{\pi}
        \Ave{\sqrt{1-(\bu\cdot\bu')^2}}\right]^{-2},
\end{equation}
where $D_{\rm r}$, the rotational diffusion coefficient of a rod 
in an isotropic, semi--dilute solution of identical rods, is related to the 
rotational diffusion constant for a dilute solution, $D_{\rm r0}$, by 
$D_{\rm r}=cD_{\rm r0}(n_{\rm p} \ell)^{-2}$ with an empirical coefficient 
$c$. 
It is generally believed that the self--consistent averaging approximation 
(\ref{Dr_approx}) does not alter the dynamics 
(\ref{Doi_kin}) qualitatively \cite{LaOe91}.

The nonlinearity of the kinetic equation (\ref{Doi_kin}) in $\psi$ 
reflects the mean--field character of the model 
that provides an effective one--body description of the many--rod 
system. 
The equilibrium distribution, $\psi^{\rm eq}$, corresponding to 
the stationary solution to the kinetic equation (\ref{Doi_kin}) 
in the absence of flow, $\bkappa=\bzero$, is given by 
$\psi^{\rm eq}\propto\exp[-U]$. The dimensionless 
self--consistent potential $U$ is identified with 
\begin{equation} \label{Uscf}
        U(\bu;\psi) = \frac{\delta F_1[\psi]}{\delta\psi(\bu)}
        = \nu\int\!d\bu' \beta(\bu,\bu')\psi(\bu'), 
\end{equation}
where Eq.~(\ref{F1_def}) has been used. 

If only excluded--volume interactions are considered, the dimensionless 
second virial coefficient $\beta(\bu,\bu')$ for rigid rods was 
obtained by Onsager \cite{On49}, 
\begin{equation} \label{2nd_virial}
        \beta(\bu,\bu') = | \bu\times\bu' | = 
        \sqrt{1-(\bu\cdot\bu')^2}. 
\end{equation}
Eq.~(\ref{2nd_virial}) has a simple geometric interpretation since 
$| \bu\times\bu' |$ is the area spanned by the vectors $\bu$ and $\bu'$. 
The excluded--volume of two cylinders with length $\ell$, diameter 
$d$ and orientations $\bu$ and $\bu'$ is therefore given by 
$d\ell^2| \bu\times\bu' |$. 
Thus, $F_1$ decreases as the polymers orient in the same direction 
and finally leads to the nematic phase when the effect of the 
excluded--volume becomes sufficiently strong. 
Note, that the second virial approximation becomes exact in the limit 
of high aspect ratio, $\ell/d\to\infty$. 

A variety of further expressions for $F_1$ have been suggested in the 
literature \cite{dGE74}. 
A particular simple, phenomenological expression was given by Maier 
and Saupe \cite{MaSa58,MaSa59}, 
\begin{equation} \label{betaMS}
        \beta^{\rm MS}(\bu,\bu') = 
        c_0 - c_1 [ (\bu\cdot\bu')^2 - \frac{1}{3} ],
\end{equation}
which together with Eq.~(\ref{F1_def}) leads to 
\begin{equation} \label{MaierS}
        F^{\rm MS}_1(\OP)=\frac{\nu}{2}[c_0 - c_1 \OP\colon\OP],
\end{equation}
where $c_0$ and $c_1$ are parameters independent of $\psi$. 
In Eq.~(\ref{MaierS}), the important notion of 
the orientational order parameter $\OP$ is introduced, 
$\OP=\ave{\bu\bu-\frac{1}{3}\bone}$. 
The isotropic state is characterized by $\OP=\bzero$, 
while $\OP\neq\bzero$ indicates (nematic) orientational ordering.  

The self--consistent equilibrium distribution corresponding to the potential 
(\ref{Uscf}) 
determines the isotropic nematic transition in equilibrium. 
The detailed form of the interaction potential can 
have significant effect on the behavior of the order parameter in the 
nematic phase. Specifically, the amount of order at 
the transition is known to be much smaller in the Maier--Saupe 
theory than in the Onsager model \cite{St73}. 

In the sequel, we consider generic interaction functionals of the form 
\begin{equation} \label{F1_n}
        F_1[\psi] = \bar{F}_1(\setM), 
\end{equation}
where $\setM=\{M_1,\ldots,M_{m_n}\}=\{\A{2},\ldots,\A{2n}\}$
denotes the set of 
irreducible (anisotropic) moments of the distribution function up to 
order $2n$, 
$(\A{2n})_{\alpha_1\beta_1\ldots\alpha_n\beta_n}=
\ave{\stl{u_{\alpha_1}u_{\beta_1}\ldots u_{\alpha_n}u_{\beta_n}}}$, 
and $\stl{\bB}$ denotes the symmetric irreducible part of the 
tensor $\bB$ \cite{HessKoehler80}. 
In other words, the functional dependence of the interaction 
functional on the orientational distribution function $\psi$ 
comes only through the dependence of the moments of $\psi$. 
For example, the tensor $\A{2}$ contains five independent components. 
For convenience, these components are denoted by the scalars 
$M_1,\ldots,M_5$. 
The total number of components of irreducible moments up to 
order $2n$ is $m_n$. In case $n=1$ one has $m_n=5$.  
In case $n>1$, the components $M_i$ for $i>5$ contain the independent 
components of the tensors $\A{2k}$, $k=2,\ldots,n$. 

The interaction functionals (\ref{F1_n}) correspond to virial 
coefficients of the form 
$\beta(\bu,\bu')=\beta((\bu\cdot\bu')^2-\frac{1}{3})$ with 
polynomials $\beta(x)$ of degree $n$. 
The Maier--Saupe expression (\ref{MaierS}) is of the form 
(\ref{F1_n}) with $n=1$. 
The Onsager potential (\ref{2nd_virial}), on the contrary, can only be 
cast into the form (\ref{F1_n}) for $n\to\infty$. 
In this case, systematic approximations of the form (\ref{F1_n}) with 
finite $n$ have been proposed in \cite{IKOe99}. 
The lowest order approximation to the free energy functional 
(\ref{F1_def}) 
obtained in \cite{IKOe99} for the Onsager potential is 
\begin{equation}
        \bar{F}_1(\OP) = 
        \frac{\nu}{\sqrt{6}}\sqrt{1-\frac{3}{2}\OP\colon\OP}.
\end{equation}
Some evidence was provided in \cite{IKOe99} that 
the lowest order approximation represents a good 
approximation to $F_1$ in case of the Onsager potential 
(\ref{2nd_virial}), at least on a representative subset 
of distribution functions.

Since the rotational diffusivity (\ref{Dr_approx}) is related to the Onsager 
interaction functional $F_1$, the same approximation can be employed 
to give \cite{IKOe99}
\begin{equation} \label{Dr_n}
         \overline{D}_{\rm r} = \overline{D}^{(n)}_{\rm r}(\setM).
\end{equation}
Inserting the lowest order approximation for $\bar{F}_1$ one obtains 
\cite{IKOe99}
\begin{equation} \label{Dr_P}
        \overline D_{\rm r}^{(1)}=
        (3\pi^2/32)D_{\rm r}[1-(3/2)\OP\colon\OP]^{-1}.
\end{equation}
The diffusion coefficient $\overline D_{\rm r}^{(1)}$ is  
positive in the entire physically meaningful range of the order 
parameter $\OP$. 
Expression (\ref{Dr_P}) should be compared with the Doi 
phenomenological result
\begin{equation} \label{Dr_Doi}
        \overline{D}_{\rm rD}=D_{\rm r}[1-(3/2)\OP\colon\OP]^{-2}.
\end{equation}
For highly oriented nematics the expression 
(\ref{Dr_P}) is preferred compared to Doi's result 
\cite{LA88}. 
It should be stressed that our derivation of 
the diffusion coefficient does not need any further assumptions or 
adjustable parameters, while the derivation of Eq.~(\ref{Dr_Doi}) 
requires the matching of $\overline D_{\rm r}$, 
resp.~$F^{\rm int, MS}$ in both, the isotropic and the fully ordered state 
\cite{Doi81,DoiEd86}.
Since the rotational diffusivity $\overline D_{\rm r}$ is only 
determined within a phenomenological constant, 
the numerical factor $3\pi^2/32$ 
in Eq.~(\ref{Dr_P}) can safely be put equal to one.

\subsection{H--Theorem}
The time evolution of the free energy functional $F=F_0+F_1$ is given by 
\begin{equation} \label{dtF_def}
        \dot{F} = \int\!d\bu\, \frac{\delta F[\psi]}{\delta\psi(\bu)} 
        \partial_t\psi(\bu). 
\end{equation}
Inserting the kinetic equation (\ref{Doi_kin}) into Eq.~(\ref{dtF_def}) 
one obtains 
\begin{equation} \label{dtF}
        \dot{F}=\bkappa\colon\btau^{\rm p} - 
        \ave{\cRu\frac{\delta F[\psi]}{\delta\psi(\bu)}\cdot 
          \bar{D}_{\rm r}\cRu\frac{\delta F[\psi]}{\delta\psi(\bu)}}. 
\end{equation}
In the absence of a velocity gradient, the dynamics (\ref{Doi_kin}) 
drives the system irreversibly to the unique equilibrium state 
$\psi^{\rm eq}$ 
with $F$ being an $H$--function of the relaxation dynamics.
In the presence of a velocity gradient, Eq.~(\ref{dtF}) also describes the 
free energy exchange between the polymer and the solvent subsystem. 
The expression (\ref{stress_LCP}) for the stress tensor 
can be rewritten in terms of partial rather then functional 
derivatives if the free energy $F$ is of the form 
$F=F_0+\bar{F}_1$, with $\bar{F}_1$ given by 
Eq.~(\ref{F1_n}) depends on $\psi$ only through moments, 
\begin{equation} \label{stress_n}
        \btau^{\rm p} 
        = 3n_{\rm p}\kb T\OP + 
        \sum_{k=1}^n \ave{[\bu\times\cRu 
          (\stl{u_{\alpha_1}u_{\beta_1}\cdots u_{\alpha_k}u_{\beta_k}})]\bu}
        \frac{\partial \bar{F}_1}
        {\partial (\A{2k})_{\alpha_1\beta_1\cdots\alpha_k\beta_k}}.
\end{equation}
In the special case $n=1$, Eq.~(\ref{stress_n}) simplifies to 
\begin{equation} \label{stress_n1}
        \btau^{\rm p} 
        = 3n_{\rm p}\kb T\OP  
        - 2\frac{\partial\bar{F}_1}{\partial\A{2}}\cdot\ba_2 + 
        2\frac{\partial\bar{F}_1}{\partial\A{2}}\colon\ba_4. 
\end{equation}
For the Maier--Saupe potential, Eq.~(\ref{MaierS}), Eq.~(\ref{stress_n1}) 
coincides with the standard expression for the stress tensor 
given in \cite{DoiEd86}.

\section{Canonical Distribution Functions in Polymer Dynamics} \label{QEA}
Within the kinetic models described in Sec.~\ref{models}, the 
dynamics and viscoelastic properties of liquid--crystalline polymers are 
given by Eqs.~(\ref{Doi_kin}) and (\ref{stress_LCP}).  
In viscoelastic flow calculations, the combined simulation of the 
Navier--Stokes equations coupled with the kinetic equation
(\ref{Doi_kin}) 
is 
computationally very intensive. 
Therefore, there has been considerable interest in deriving 
sufficiently accurate closed form 
constitutive equations from kinetic theory. 

\subsection{The Closure Problem}
Isotropic, uniaxially and biaxially oriented 
phases of LCPs are 
conveniently described by the order parameter tensor $\OP$. 
It is reasonable therefore to look for closed descriptions of the 
dynamics in terms of $\OP$ alone. 
The description of the viscoelastic properties of LCPs in terms of 
the stress tensor (\ref{stress_n}), however, requires also the knowledge 
of moments up to order $2n+2$. 
The time evolution of moments $\A{2j}$ obtained from the kinetic 
equation (\ref{Doi_kin}) is of the form 
\begin{equation} \label{dtA_j}
        \frac{d}{dt}\A{2j} = 
        \bG^{(2j)}(\A{2},\ldots,\A{2j+2n})
\end{equation}
with the functions  
\begin{eqnarray} \label{G_j}
        \lefteqn{ G^{(2j)}_{\alpha_1\beta_1\cdots\alpha_j\beta_j}
        (\A{2},\ldots,\A{2j+2n}) = 
        \ave{\cRu[\stl{u_{\alpha_1}u_{\beta_1}\cdots u_{\alpha_j}u_{\beta_j}}]
          \cdot(\bu\times\bkappa\cdot\bu)} }\nonumber\\
        &&{}+ \bar{D}^{(n)}_{\rm r}
        \ave{\cRu^2[\stl{u_{\alpha_1}u_{\beta_1}\cdots u_{\alpha_j}u_{\beta_j}}]}
        \nonumber\\
        &&+ \bar{D}^{(n)}_{\rm r} \sum_{k=1}^n 
        \ave{\cRu[\stl{u_{\alpha_1}u_{\beta_1}\cdots u_{\alpha_j}u_{\beta_j}}]
          \cdot\cRu[\stl{u_{\alpha_1}u_{\beta_1}\cdots 
          u_{\alpha_k}u_{\beta_k}}]}\frac{\partial\bar{F_1}}
        {\partial (\A{2k})_{\alpha_1\beta_1\cdots\alpha_k\beta_k}}. 
\end{eqnarray}
Eqs.~(\ref{dtA_j}) and (\ref{G_j}) form a hierarchy of moment equations 
where moments of order $2j$ couple to moments of order $(2j+2n)$. 
Therefore, closed form constitutive equations in terms of low order moments 
cannot be derived exactly from the kinetic equation but require some 
approximations of the distribution function. 
There exist an enormous amount of closure approximations 
as well as many numerical tests of the approximations in various 
flow situations
(see, e.g.~\cite{Doi81,FeChLe98,ChLeFr95,ChLe98,Wa97a,Wa97b}).

\subsection{Extremum Principle and Canonical Distribution Functions}
Following the approach described in \cite{IKO02}, 
we here derive closed form constitutive equations for LCPs by applying the 
so--called quasi--equilibrium approximation. 
Within this approximation, canonical distribution functions are obtained 
from the extremum principle under constraints \cite{BALIAN92,GORBAN84}
\begin{equation} \label{MaxEnt}
        F[\psi]\to {\rm min},\quad 
        1=\int\!d\bu\, \psi(\bu),\ \ 
        M_k=\int\!d\bu\, m_k(\bu)\psi(\bu)
\end{equation}
for $k=1,\ldots,m_n$. 
Notice, that this principle does not coincide with the entropy maximum 
principle, because the free energy $F$, in general, 
is not proportional to the entropy. 
We consider in the sequel the special choice 
$\stl{u_{\alpha_1}u_{\beta_1}\cdots u_{\alpha_k}u_{\beta_k}}$ 
for the functions $m_k(\bu)$, 
such that $\setM$ represents the set of the first $n$ irreducible 
moments of the distribution function 
already introduced in Eq.~(\ref{F1_n}). 
Inserting the functional form $F=F_0+\bar{F}_1$, 
where $F_0$ is given by Eq.~(\ref{F0_def}) 
and $\bar{F}_1$ by Eq.~(\ref{F1_n}), 
the solution to Eq.~(\ref{MaxEnt}) is given explicitly as 
\begin{equation} \label{fQE}
        \QE(\bu) = \psi^{\rm eq}(\bu)\exp[\sum_{k=1}^{m_n} 
        \Lambda_k m_k(\bu) +
        \Lambda_{0}].
\end{equation} 
The set of Lagrange multipliers 
$\setL=\{\Lambda_1,\ldots,\Lambda_{m_n}\}=\{\La{2},\ldots,\La{2n}\}$ 
ensure the constraints in Eq.~(\ref{MaxEnt}), 
\begin{equation} \label{A_Lambda}
        M_k = \int\!d\bu\, m_k(\bu)
        \psi^{\rm eq}(\bu)\exp[\sum_{j=1}^{m_n} 
        \Lambda_j m_j(\bu)+\Lambda_{0}],
\end{equation} 
and $\Lambda_{0}$ ensures the normalization of $\QE$. 
The interpretation of the CDF (\ref{fQE}) as a
quasi--equilibrium distributions has already been given in \cite{Gu98}. 

The quasi--equilibrium free energy $F^\ast$ is defined as 
$F^\ast=F[\QE]$. Inserting the form (\ref{fQE}) one obtains 
\begin{equation} \label{F_macro}
        F^\ast(\setM) = 
        \sum_{k=1}^{m_n} \Lambda_k M_k + \Lambda_{0}.
\end{equation}
Note, that although the interaction part of the free energy 
is given by Eq.~(\ref{F1_n}), the total quasi--equilibrium 
free energy $F^\ast$ cannot be expressed as a function of 
$\setM$ explicitly. This difficulty results from the 
fact that the normalization $\Lambda_{0}$ of the 
distribution (\ref{fQE}) cannot be evaluated analytically in general. 
From Eq.~(\ref{F_macro}), the Lagrange multipliers can be interpreted 
as the conjugate to the macroscopic variables, 
\begin{equation} \label{lambda_conj}
        \Lambda_k = \frac{\partial F^\ast(\setM)}{\partial M_k}, 
        \quad k=1,\ldots,m_n.
\end{equation}

\subsection{Macroscopic dynamics} \label{macro_dyn}
In the following, the canonical distribution functions 
(\ref{fQE}) are employed in order to derive the macroscopic description of 
the polymer dynamics. 
We assume, that the CDF can be considered as representative states in the 
sense that a set of moments of the distribution function is approximated 
accurately by the corresponding moments evaluated with the CDF. 
A measure of the accuracy of this approximation is presented in 
Sec.~\ref{defect}. For improvements on this approximation see 
\cite{GoKa94,GoKaIlOe01}. 

The macroscopic time evolution is defined as the time evolution 
of the macroscopic variables evaluated with the CDF, 
$\dot{M}_k=\int\!d\bu\, m_k(\bu)\, \partial_t\QE$. 
Inserting the kinetic equation (\ref{Doi_kin}) for $\partial_t\psi$, 
one finds 
\begin{equation} \label{dtA_QE}
        \dot{M}_k = \bar{G}_k(\setM)
\end{equation}
with 
\begin{equation} \label{G_QE}
        \bar{G}_k(\setM)=
        \aveqe{[\cRu\, m_k(\bu)]
          \cdot(\bu\times\bkappa\cdot\bu)} + \sum_{j=1}^{m_n} 
        M^\ast_{kj}\Lambda_j,
\end{equation}
where the symmetric, positive semi--definite matrix $\bM$ is defined by 
\begin{equation} \label{M_QE}
        M^\ast_{kj} 
        = \bar{D}^{(n)}_{\rm r} 
        \aveqe{[\cRu\, m_k(\bu)]
          \cdot[\cRu\, m_j(\bu)]}
\end{equation}
Eqs.~(\ref{dtA_QE}), (\ref{G_QE}), (\ref{M_QE}) 
together with Eq.~(\ref{fQE}) represent the 
closed set of macroscopic equations. 
The constitutive relation is obtained by evaluating the stress tensor 
(\ref{stress_LCP}) with the CDF, 
$\btau^{\rm p,\ast}(\setM)=\btau^{\rm p}[\QE]$. 
Here, it reads 
\begin{equation} \label{stress_QE}
        \btau^{\rm p,\ast}(\setM) = 
        -3n_{\rm p}\kb T \sum_{k=1}^{m_n}
        \aveqe{[\bu\times\cRu\, m_k(\bu)]\bu}
        \frac{\partial F^\ast(\setM)}{\partial M_k}
\end{equation}
The macroscopic free energy change, $\dot{F}^\ast$, is found from 
Eqs.~(\ref{dtF}) and (\ref{dtA_QE}) to be given by
\begin{equation} \label{dtF_QE}
        \dot{F}^\ast = \bkappa\colon\btau^{\rm p,\ast} - 
        \sum_{k,j=1}^{m_n} \Lambda_k M^\ast_{kj} \Lambda_j
\end{equation}
and could have been obtained also by evaluating the free energy change 
(\ref{dtF}) on the CDF (\ref{fQE}). 
In the absence of flow, the free energy change (\ref{dtF_QE}) 
is negative semi--definite as is the underlying kinetic model.

\subsection{Alignment Tensor Models} \label{Peterlin} 
So far, the macroscopic description of LCP considered in the literature 
has almost exclusively considered the order parameter tensor $\OP$ as 
the only macroscopic variables. 
In our notation, this corresponds to the special choice 
of the order parameter tensor $\OP$ 
as the only macroscopic variable, $n=1$ and 
$\A{2}=\ave{\stl{\bu\bu}}$. 
Due to symmetry, $\A{2}$ is the lowest non--trivial moment 
of $\psi$. 

In this approximation, the dimensionless 
mean--field interaction potential derived from 
the free energy functional in case of the Onsager potential reads 
\cite{IKOe99}, 
\begin{equation} \label{V_OnsP}
        U^{(1)}(\bu,\OP) = 
        \frac{\nu}{\sqrt{6}} 
        \frac{1-\frac{3}{2}\bu\bu\colon\OP}
        {\sqrt{1 - \frac{3}{2}\OP\colon\OP}},
\end{equation}
which can be compared to the expression for the Maier--Saupe potential 
\begin{equation} \label{V_MaierS}
        U_{\rm MS}(\bu,\OP) = c_2 - c_1\nu \bu\bu\colon\OP,
\end{equation}
where $c_2$ is an arbitrary constant.

In this approximation, the canonical distribution 
function (\ref{fQE}) reduces to 
\begin{equation} \label{QEBingham}
        \psi^\ast_\Lambda(\bu) = 
        \exp{[-\nu\beta'(\OP\colon\OP)\OP+\La{2})\colon\bu\bu 
        + \Lambda_0]}/
        Z^{(1)}. 
\end{equation}
The Lagrange multipliers $\La{2}$ form 
a symmetric traceless matrix dual to $\OP$. 
The quasi--equilibrium distribution (\ref{QEBingham}) 
not only reduces to the equilibrium distribution for 
$\La{2}=\bzero$ but also gives the exact steady state solution 
$\La{2}^{\rm ss}=-\dgamma/2$ for homogeneous potential flows. 
Due to the occurrence of the second moment $\A{2}$, 
the quasi--equilibrium states $\psi^\ast_\Lambda$ have 
to be determined self--consistently. 
Further manipulations are simplified by introducing new 
variables 
$\bTheta = \La{2} - \nu \beta'(\OP\colon\OP)\OP$ and 
$\Theta_0= \Lambda_0 - \ln Z^{(1)}$, 
where $\bTheta$ is a symmetric, traceless matrix. 
In terms of new variables, 
the quasi--equilibrium states (\ref{QEBingham}) take the form 
\begin{equation} \label{Bingham}
        \psiQE(\bu) = 
        \exp{[\bu\cdot\bTheta\cdot\bu + \Theta_0]}, 
\end{equation}
where $\Theta_0$ ensures the normalization of 
$\psiQE$.
The distribution (\ref{Bingham}) is sometimes termed 
Bingham distribution.

Accordingly, the diffusion coefficient $\overline D_{\rm r}$ 
is approximated by $\overline D_{\rm r}^{(1)}$ given in Eq.~(\ref{Dr_P}). 
In the following, 
the macroscopic time evolution equations for $\A{2}$ are obtained 
from Eqs.~(\ref{dtA_QE}), (\ref{G_QE}), (\ref{M_QE}) and read,   
\begin{equation} \label{dtA2}
        \dot{\OP} = \bG^\ast_h(\OP) + \bG^\ast_d(\OP). 
\end{equation}
The presence of a flow field gives raise to the contribution $\bG^\ast_h$ 
in Eq.~(\ref{dtA2}), 
\begin{equation} \label{dtA2_G}
        \bG^\ast_h(\OP) = \bkappa\cdot\OP + \OP\cdot\bkappa^{\rm T} 
        +\frac{1}{3} \dgamma 
        -\dgamma \colon \Afourqe ,
\end{equation}
where $\dgamma=\bkappa+\bkappa^{\rm T}$ denotes the 
rate--of--strain tensor and 
$\Afourqe=\aveqe{\bu\bu\bu\bu}$ is evaluated with the Bingham 
distribution (\ref{Bingham}). 
The contribution $\bG^\ast_d$ of the Brownian motion and the interaction 
potential is given by 
\begin{equation} \label{dtA2_GEN}
         \bG^\ast_d(\OP) = \bM^\ast\colon\La{2},  
\end{equation}
where $\La{2}=\partial F^\ast(\OP)/\partial \OP$ and 
the matrix $\bM^\ast$ is given by 
\begin{eqnarray} \label{M2_AA_LCP}
        M^\ast_{\alpha\beta\mu\nu} & = &  
        \overline D_{\rm r}(
        \delta_{\alpha\mu}(\A{2})_{\beta\nu} + 
        \delta_{\alpha\nu}(\A{2})_{\beta\mu} + 
        \delta_{\beta\mu}(\A{2})_{\alpha\nu} + 
        \delta_{\beta\nu}(\A{2})_{\alpha\mu} - 
        4 (\ba^\ast_4)_{\alpha\beta\mu\nu}  ). \nonumber
\end{eqnarray}
In Eq.~(\ref{M2_AA_LCP}), use has been made of the fact that the 
approximated diffusion coefficient $\overline D_{\rm r}$ is independent 
of the orientation $\bu$. 
Inserting the expression for the free energy one arrives at
\begin{equation} \label{dtA2_F}
        \bG^\ast_d(\OP) = -6\overline D_{\rm r}^{(1)} \OP +  
        6\overline D_{\rm r}^{(1)} \nu\beta'(\OP\colon\OP)
        (\OP\cdot \OP + \frac{1}{3}\OP 
        - \OP \colon \Afourqe), 
\end{equation}
The Bingham distribution (\ref{Bingham}) has already been employed 
in the literature \cite{ChLe98,GrMaDu00} 
to derive closed form constitutive equations from 
the kinetic model (\ref{Doi_kin}). 
However, the expression for $\bG^\ast_d$ differs from the results 
given in \cite{ChLe98,GrMaDu00} not only in the 
diffusion coefficient $\overline D_{\rm r}^{(1)}$. 
It also generalizes earlier result to general interaction potentials 
of the form (\ref{F1_n}). 
For the approximation (\ref{V_OnsP}) to the Onsager potential, 
for example, a non--polynomial dependence on the order parameter 
$\OP$ occurs, which becomes important in the nematic state 
\cite{IKOe99}.
Comparative studies of various closure approximations to the 
kinetic model for the Maier--Saupe potential have clearly demonstrated 
that closures based on the Bingham distribution are preferable 
compared to more ad hoc closure approximations 
\cite{FeChLe98}.

In the present case, the macroscopic free energy (\ref{F_macro}) 
reduces to 
\begin{equation} \label{S_QE_A2}
        F^\ast(\OP) = \La{2}\colon\A{2} + \Lambda_0.
\end{equation}
The macroscopic polymer contribution to the stress tensor (\ref{stress_QE}) 
simplifies to 
\begin{equation} \label{stress_LCPmacro}
        \btau^{\rm p,\ast} = 2\kb T \left( 
        \OP\cdot\La{2} - \Afourqe\colon\La{2} \right).
\end{equation}
Expression (\ref{stress_LCPmacro}) is also obtained when the 
macroscopic entropy (\ref{S_QE_A2}) is inserted into 
Eq.~(\ref{stress_LCP}).

\section{Numerical Integration Scheme} \label{DualInt_Chapt}
Eqs.~(\ref{dtA_QE}) are the closed form macroscopic equations. 
Together with the expression (\ref{stress_QE}) 
for the stress tensor they provide 
the macroscopic constitutive equation of the system. 
However, the time evolution equations (\ref{dtA_QE}) still contain the 
Lagrange multipliers, that have to be determined such that the 
constraints in Eq.~(\ref{MaxEnt}) are satisfied. 

To deal with this situation, 
a numerical integration scheme for the macroscopic equations 
in the context of dilute solutions of flexible polymers 
has been proposed in Ref.~\cite{IKO02}. 
This scheme is general enough to be applied also to the present situation. 
We present here this integration scheme only briefly, referring 
the reader to \cite{IKO02} for further details.

\subsection{Dynamics of Dual Variables} \label{DualDyn}
In the extremum principle (\ref{MaxEnt}), the Lagrange multipliers 
occur naturally to satisfy the constraints. 
Due to Eq.~(\ref{lambda_conj}), 
the Lagrange multipliers are interpreted as conjugate variables. 
Instead of eliminating the Lagrange multipliers $\setL$ in favor of the 
macroscopic variables $\setL=\setL(\setM)$, the Lagrange multipliers 
play the role as dynamical variables 
that determine the values of the macroscopic 
variables, $\setM=\setM(\setL)$. 
Note, that the functions $\setM(\setL)$ are given by Eq.~(\ref{A_Lambda}).    
With the help of the Legendre transform of the macroscopic free energy, 
$G(\setL)=F^\ast(\setM)+\sum_{k=1}^{m_n}\Lambda_kM_k$, 
the time evolution of $\setL$ is given by 
\begin{equation} \label{dt_Lambda}
        \dot{\Lambda}_k = \sum_{j=1}^{m_n}(\bC)^{-1}_{kj}\dot{M}_j(\setL),
\end{equation}
where $\bC^{-1}$ is the inverse of the matrix 
$C_{kj}=\partial^2 G(\setL)/\partial\Lambda_k\partial\Lambda_j$ and 
$\dot{M}_j(\setL)$ is given by the right hand side of 
Eq.~(\ref{dtA_QE}).

\subsection{Numerical Integration Scheme}
\label{DualInt}
The reformulation of the macroscopic dynamics (\ref{dtA_QE}) 
in terms of the dual variables $\setL$ described in Sec.~\ref{DualDyn} 
is suitable for numerical implementation. 
The Lagrange multipliers $\setL$ now play the role of independent 
dynamic variables, instead of $\setM$. 
In order to advance given values $\setL(t)$ at time $t$ to their values 
$\setL(t+\tau)$, with small time step $\tau$, the following first order 
integration scheme is proposed in Ref.~\cite{IKO02}: 
\begin{enumerate}
\item The new values of the macroscopic variables $\setM(t+\tau)$ 
are found from the values $\setL(t)$ by 
\begin{equation} \label{dualInt1}
  \frac{M_k(t+\tau)-M_k(t)}{\tau} = 
  G_k(\setM(\setL(t))),
\end{equation}
where $G_k(\setM)$ denotes the right hand side of 
Eq.~(\ref{dtA_QE}) 
with $\setL(t)$ the actual values of the Lagrange multipliers. 
\item The matrix $C_{kj}$ is evaluated from 
\begin{equation} \label{dualInt2}
  C_{kj}(t) = \aveqe{m_km_j}(t) -M_k(t)M_j(t),
\end{equation}
where $\aveqe{m_km_l}$ is calculated with the distribution function 
(\ref{fQE}) where $\setL=\setL(t)$. 
\item The $n\times n$ matrix $\bC(t)$ is inverted numerically to 
give $C^{-1}(t)$.   
\item The values of the Lagrange multipliers, $\setL(t+\tau)$, 
are given by 
\begin{equation} \label{dualInt3}
  \frac{\Lambda_k(t+\tau)-\Lambda_k(t)}{\tau} = 
  \sum_{j=1}^{m_n}C^{-1}_{kj}(t) \frac{M_j(t+\tau)-M_j(t)}{\tau}
\end{equation}
\end{enumerate}
This concludes one time step of integration. 
The integration scheme has to be supplemented by initial conditions 
$\setL(0)$. The special case of equilibrium initial conditions 
corresponds to $\setL(0)=\underline{0}$. 
The appearance of the correlation matrix $\bC$ as a tool to 
recompute moments into Lagrange multipliers in this scheme 
is intimately related to the structure of the quasi--equilibrium 
approximation. 

Note, that evaluating the matrix $\bC$ requires moments of 
$\Psi$ which are of higher order than the macroscopic variables 
themselves. If $n$ denotes the number of macroscopic variables, 
the numerical integration scheme requires in total $n(n+3)/2$ 
integrals per time step to evaluate $\setM$ and $\bC$. 
Due to the symmetry of $\bC$, this number is of order $n^2/2$ rather 
than $n^2$. 
The evaluation of all these integrals with standard numerical 
methods might be time--consuming, 
especially for high--dimensional integrals. 
It is demonstrated in \cite{HuBa98} that under certain circumstances 
these integrals can be evaluated
efficiently by adapting methods of fast Fourier transformations.
This computational issue is out of scope of this paper.  

\subsection{Numerical Implementation of Alignment Tensor Models}
For the special case of alignment tensor models described in 
Sec.~\ref{Peterlin}, the general integration scheme outlined 
in Sec.~\ref{DualInt} is applied. 
The integrals over the unit sphere 
associated with the evaluation of moments of the 
canonical distribution function are performed numerically. 
Different resolutions of the numerical integration are used to 
obtain more accurate results by extrapolating to infinite resolution. 
The normalization of $\bu$, $\bu^2=1$, and the consistency 
relations 
$\aveqe{\bu\bu\bu\bu}\colon\bone=\aveqe{\bu\bu}$
are used as additional checks of the results of the numerical 
integration. 
The inversion of the matrix $\bC$ is done by using the explicit 
expression for the inverse of a symmetric $6\times 6$ matrix 
as given by the symbolic computation software 
Mathematica$^{\rm TM}$. 
The Lagrange multipliers are advanced by Eq.~(\ref{dualInt3}). 
The resulting algorithm has been implemented and tested under 
various circumstances. 
This numerical implementation differs from the one 
proposed in \cite{ChLe98}. 
In the latter reference, approximate expressions for the 
eigenvalues of $\ba^\ast_4$ as a function of the eigenvalues 
of $\OP$ are proposed, 
whereas in the present approach the necessity of expressing the 
Lagrange multipliers $\setL$ in terms of the moments is 
circumvented by the scheme (\ref{dualInt1})--(\ref{dualInt3}). 
Together with suitable rotations to diagonalizing coordinate 
systems, these expressions allow direct numerical integration 
of Eq.~(\ref{dtA2}).

\section{Accuracy of the Approximation by CDF} \label{defect}
The use of the CDFs for the macroscopic description described 
in Sec.~\ref{macro_dyn} imposes 
approximations to the underlying dynamics. 
Therefore, the accuracy of this approximation needs to be 
discussed. 

The accuracy of the approximation by CDFs can be studied in 
two ways: First, the result of integrating the macroscopic dynamics 
can be compared to direct numerical integration (e.g.~by 
Brownian dynamics) of the underlying kinetic model. 
Alternatively, the dynamic variance $\Delta$ that plays a 
central role in the method of invariant manifold \cite{GoKa94} 
may be used in order to access the accuracy of the
approximation. 
The advantage of using $\Delta$ to estimate the accuracy of the 
approximation is that it can be evaluated while integrating the 
macroscopic dynamics, {\it without} performing numerics for the 
underlying microscopic dynamics. 
For the use of the dynamic variance in a mixed micro--macro 
computation see \cite{GoKaIlOe01}. 

The dynamic variance $\Delta$ is defined as 
\begin{equation} \label{delta_def}
        \Delta(\bu;\setM) = J\QE - 
        \sum_{j=1}^{m_n}\frac{\partial\QE}{\partial M_j}\dot{M}_j^\ast,
\end{equation}
where $J\QE$ denotes the right hand side of Eq.~(\ref{Doi_kin}) evaluated 
with the CDF $\QE$, 
\begin{equation} \label{J_QE}
        J\QE = -\cRu\cdot\left[ 
        \bu\times\left( \bkappa\cdot\bu\QE \right) \right] 
      + \overline{D}_r 
        \sum_k^{m_n}\cRu\cdot\QE\cRu m_k(\bu)\Lambda_k.
\end{equation}
The dynamic variance $\Delta$ can be interpreted as the difference 
of the microscopic time evolution evaluated with $\QE$, 
Eq.~(\ref{J_QE}), and the time evolution of the CDFs due to the 
macroscopic dynamics. 
If $\Delta\equiv 0$, the CDFs are said to form an invariant manifold 
of the dynamics (\ref{Doi_kin}). 
In case $\Delta\neq 0$, the method of invariant manifold uses 
$\Delta$ to obtain improved manifolds of distribution functions. 
Here and in a companion paper \cite{IKO02}, $|\Delta|$ is used as 
a measure of the accuracy of the approximation by CDFs, where the 
norm $|\bullet |$ is specified later. 

Consider functions $Y_k(\setM)$ defined by 
$Y_k(\setM)\equiv\int\!d\bu\, y_k(\bu)\Delta(\bu;\setM)$. 
Some choices of $y_k(\bu)$ are discussed below. 
The functions $Y_k$ can be interpreted as the difference of the 
time evolution of the macroscopic moments if evaluated from the 
microscopic dynamics and from the macroscopic dynamics. 
For the special case of alignment tensor models, introduced in
Sec.~\ref{Peterlin}, functions 
$Y_k$ found from Eq.~(\ref{delta_def}) read 
\begin{eqnarray}
        Y_k & = & \aveqe{y_k(\bu) \bu\bu\colon[
          3\bkappa-2\bTheta\cdot\bP\cdot\bkappa + 
          4\DD\bTheta\cdot\bP\cdot\La{2}-6\DD\La{2}
          ]} \nonumber\\
        &&{}
        - [\aveqe{y_k(\bu)(\bu\bu)}-\aveqe{y_k(\bu)} 
          \aveqe{\bu\bu}]\colon\frac{\partial \bTheta}{\partial \OP} 
        \colon\dot{\OP},
\end{eqnarray}
where $\bP=\bone-\bu\bu$ is the projector perpendicular to $\bu$, 
$\dot{\OP}$ is given by Eq.~(\ref{dtA2}) and 
\begin{equation}
        \frac{\partial \bTheta}{\partial \OP} = 
        \bC^{-1} - \nu\beta' \bone\bone - 2\nu\beta''\OP\OP.
\end{equation}
We have $Y_0=0$ for $y_0=1$, since the first and second term in 
Eq.~(\ref{delta_def}) conserve the normalization of the distribution 
function independently. 
Further, $Y_k$ vanishes by construction if $y_k$ corresponds to 
components of the tensor $\bu\bu$. 
For $y_k$ corresponding to $\bu\bu\bu\bu$ or higher order tensors, 
the quantities $Y_k$ contain valuable information about the 
variance $\Delta$ and may be used as a measure for $|\Delta |$.

\section{Some Numerical Results} \label{dualLCP}
In the present section, 
the macroscopic equations (\ref{dtA2_GEN}) with (\ref{Bingham}) 
are integrated numerically using the dual integrator presented in 
Sec.~\ref{DualInt}.

\subsection{Some Numerical Results for Steady Shear Flow} 
\begin{figure} 
  \begin{center}
     \includegraphics[width=8cm,height=6cm]{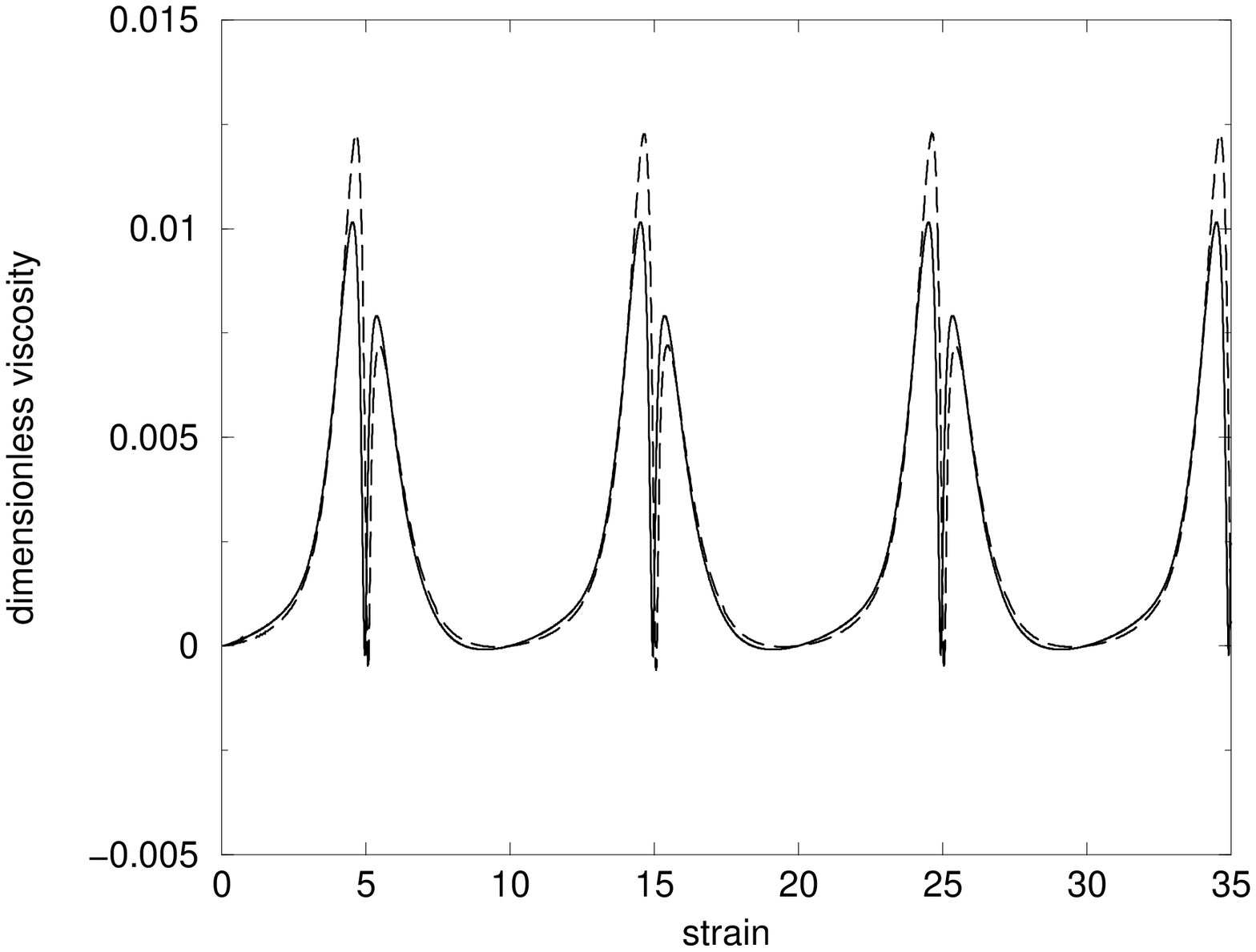}
     \caption[shear viscosity of LCPs]{ 
        \label{OnsP_dual_shear_visc} 
        Time evolution of the dimensionless shear viscosity as a 
        function of shear strain. The nematic potential was chosen 
        to be $\nu_2=6.5$. 
        Solid and dashed line correspond to the dimensionless 
        shear rate ${\rm Pe}=5$ and ${\rm Pe}=10$, respectively.}
  \end{center}
\end{figure}

The integration scheme described above was used to obtain the 
time evolution of the macroscopic variables $\A{2}$ from 
Eq.~(\ref{dtA2_GEN}) in steady shear flow. 
The functions $\bG$ and $\bF$ are given in Eqs.~(\ref{dtA2_G}) 
and (\ref{dtA2_F}), respectively with 
$\kappa_{\alpha\beta}=\dot{\gamma}\delta_{\alpha1}\delta_{\beta2}$. 
We employ the approximation 
(\ref{V_OnsP}) to the Onsager excluded volume potential and 
expression (\ref{Dr_P}) for the diffusion coefficient. 
The minimal value $\nu_2$ of the nematic potential that allows 
only a stable nematic phase was estimated from steady state 
values of the numerical solution of Eq.~(\ref{dtA2_GEN}) 
in the absence of flow, $\bG=\bzero$. 
For almost isotropic initial conditions a value $\nu_2\approx 6.5$ 
was obtained. 
Following Larson \cite{La90}, we set $\nu_2$ equal to this minimal 
value. 
Fig.~\ref{OnsP_dual_shear_visc} shows the dimensionless 
shear viscosity as a function of shear strain for two low 
values of the dimensionless shear rate P\'e 
as obtained form the numerical integration for equilibrium 
initial condition. 
The P\'eclet number P\'e is defined as the   
dimensionless shear rate
$\mbox{P\'e}=\dot{\gamma}/6D_{\rm r}^\ast$, 
where $D_{\rm r}^\ast$ is the diffusion coefficient 
$D_{\rm r}$ defined in Eq.~(\ref{Dr_approx}) and below 
at the concentration $\nu_2$. 
It is obvious from Fig.~\ref{OnsP_dual_shear_visc} that 
the present macroscopic description shows a so--called 
`tumbling' or `wagging' regime \cite{La90}. 

\begin{figure} [ht]
  \begin{center}
     \includegraphics[width=8cm,height=6cm]{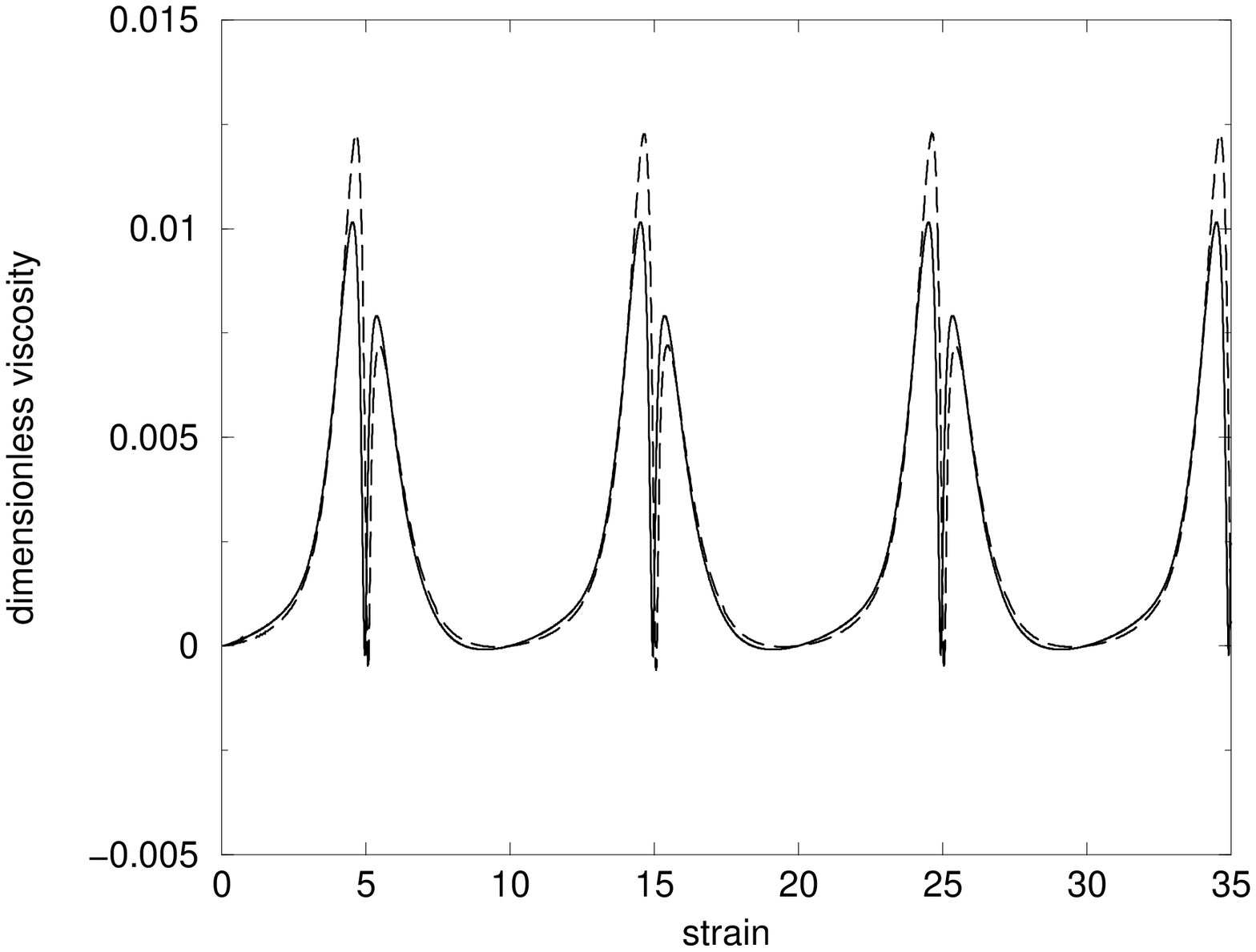}
     \caption[Experimental shear viscosity of LCP ]{ 
        \label{visc_strain_jamieson} 
        Time evolution of the shear viscosity of 
        n--octylcyanobiphenyl as a function of shear strain for 
        shear rate $\dot{\gamma}=16\, {\rm s}^{-1}$.  
        Data are taken from \cite{JGCS96}.}
  \end{center}
\end{figure}

Rheological and rheo--optical experiments on lyotropic 
solutions of rodlike polymers have clearly demonstrated the 
existence of a `tumbling' or `wagging' regime for low 
shear rates. 
In Fig.~\ref{visc_strain_jamieson}, the experimental result for  
the shear viscosity of the lyotropic nematic polymer 
n--octylcyanobiphenyl is given as reported in 
\cite{JGCS96}. 
Note, that the experimental results of Refs.~\cite{JGCS96,Jamieson98} 
are obtained from birefringence data of a monodomain 
liquid--crystalline polymer. 
Therefore, neglecting spatial inhomogeneities in the kinetic model 
is justified. 
The experimental data were obtained in a parallel--plate flow 
from birefringence measurements for a shear rate of  
$\dot{\gamma}=16\, {\rm s}^{-1}$. 
Qualitatively, the simulation results show similar behavior 
than the experimental findings. 
However, the damping of the oscillations of the shear viscosity 
is not seen in Fig.~\ref{OnsP_dual_shear_visc}. 
The authors of Ref.~\cite{JGCS96} speculate that this behavior 
might be due to out--of--plane director orientations. 
Another dissipative mechanism not included in the present kinetic 
model is translational diffusion which could also explain the 
damping of the oscillations in Fig.~\ref{OnsP_dual_shear_visc}. 
We mention, that the authors of Ref.~\cite{Jamieson98} obtained a fit to 
experimental data on a similar liquid--crystalline polymer 
which is comparable to our simulation results.  
Their procedure requires fitting of three parameters 
of a generalized version of Ericksen's theory of nematic 
liquid crystals \cite{Ericksen60}.

\section{Conclusion}
\label{end}
A systematic approach to the derivation and implementation of 
constitutive equations proposed in \cite{IKO02} is applied 
in the present work to the dynamics of liquid--crystalline 
polymers. 
Employing the quasi--equilibrium approximation, a set of 
canonical distribution functions is obtained which is further 
used to derive constitutive equations. 
The numerical implementation and a measure of the accuracy of 
the approximation is discussed. 
The present approach is illustrated for the kinetic model of 
liquid--crystalline polymers with the Onsager excluded volume 
potential in steady shear and steady elongational flow.

\section*{Acknowledgments}
This research was supported in part by the National Science
Foundation under grant No.~PHY99-07949.


\pagebreak
%
%

%
%
%

\end{document}